\pdfoutput=1
\documentclass[]{aa}

\usepackage{txfonts}
\usepackage{xcolor}
\usepackage{graphicx}
\bibpunct{(}{)}{;}{a}{}{,} % to follow the A&A style
\newcommand{\Mm}{{\mathrm{\, Mm}}}
\newcommand{\kms}{{\mathrm{\, km \; s^{-1}}}}

\newcommand{\mins}{{\mathrm{\, minutes}}}

\begin{document} 

  \title{Exploration of long-period oscillations  in an H$\alpha$ prominence}

 %  \subtitle{}

   \author{
          %\inst{1}
          M. Zapi{\' o}r \inst{1}
          \and 
         B. Schmieder \inst{2}
          \and
          P. Mein\inst{2}
          \and
          N. Mein\inst{2}
           \and
          N. Labrosse \inst{3}
                   \and
          M. Luna \inst{4,5}
          }
   \institute{Astronomical Institute ASCR, Fričova 298, 251 65 Ondřejov, Czech Republic\\
   \email{maciej.zapior@asu.cas.cz}
         \and
                LESIA, Observatoire de Paris, PSL Research University, CNRS Sorbonne Universit\'e, Univ. Paris 06, Univ. Paris Diderot, Sorbonne Paris Cit\'e, 5 place Jules Janssen, Meudon, 92195, France\\                
                                \and
                SUPA School of Physics and Astronomy, University of Glasgow,
              Glasgow, G12 8QQ, UK\\             
              \and
                Instituto de Astrof\'{\i}sica de Canarias, E-38200 La Laguna, Tenerife, Spain\\
        \and
        Departamento de Astrof\'{\i}sica, Universidad de La Laguna, E-38206 La Laguna, Tenerife, Spain}
             
%\altaffiltext{1}{Instituto de Astrof\'{\i}sica de Canarias, E-38200 La Laguna, Tenerife, Spain; \email{mluna@iac.es}}
%\altaffiltext{2}{Departamento de Astrof\'{\i}sica, Universidad de La Laguna, E-38206 La Laguna, Tenerife, Spain}
   \date{Received ...; accepted ...}

% \abstract{}{}{}{}{} 
% 5 {} token are mandatory
 
  \abstract
  % context heading (optional)
  % {} leave it empty if necessary  
   %Tornadoes are best observed with  SDO/AIA  using its 193 \AA\ filter, and appear like dark silhouettes that give the impression of rotating around their axis.
{  {  In previous work, we studied a prominence which appeared  like a tornado in  a  movie made from  193 \AA\ filtergrams  obtained with the  {Atmospheric Imaging Assembly}  (AIA) imager aboard  the \textit{Solar Dynamics Observatory} (SDO).
    The observations  in H$\alpha$  obtained simultaneously during two  consecutive sequences of one hour  with the Multi-channel Subtractive Double Pass Spectrograph} (MSDP)  operating at the solar tower in Meudon   showed that  the cool plasma inside the tornado  was  not rotating around  its vertical axis.  Furthermore, the evolution of the Dopplershift  pattern suggested the existence of oscillations of periods close to the time-span of each sequence.} 
  % aims heading (mandatory)
    {{ The aim of the present   work is  to    assemble  the two sequences of H$\alpha$ observations as a full data set  lasting  two hours to confirm the existence of oscillations,  and  determine  their nature.}
    }
  % methods heading (mandatory)
 { { 
 After having coaligned the Doppler maps of 
 %We co-aligned  
 the  two   sequences, 
 we use a Scargle periodogram analysis and cosine fitting to compute the periods and the phase  of the  oscillations in  the full data set.}
 }
  % results heading (mandatory)
   { {  Our analysis  confirms the existence of  oscillations with periods between 40 and 80 minutes.   In the   Dopplershift maps, we identify large areas  with   strong spectral power. In two of them,  the oscillations of individual pixels are in phase. However, in the top area of the prominence,   the phase is varying slowly, suggesting    wave propagation. }
   }
  % conclusions heading (optional), leave it empty if necessary 
   {{   We   conclude that the  prominence  does  not oscillate as a whole structure but exhibits different  areas with their own oscillation periods and characteristics: standing or propagating waves.  We  discuss the nature of the standing oscillations and the propagating waves. These can be interpreted in terms of gravito-acoustic modes and magnetosonic waves, respectively.}
 }


   \maketitle
%
%________________________________________________________________

\section{Introduction}

%__________________________________________________________________

{ The \textit{Solar Dynamics Observatory} \citep[\textit{SDO};][]{Pesnell2012} spacecraft, thanks to its high-spatial- and temporal-resolution  \textit{Atmospheric Imaging Assembly} imager \citep[\textit{AIA};][]{Lemen2012}, has
 evidenced the strong dynamic nature of solar prominences, for example high flows in quasi-vertical structures, rising bubbles, and rotating,  tornado-like structures  \citep{Dudik2012,Orozco2012,Wedemeyer2013,Berger2014,Su2014,Levens2015}.  
Using the Hinode Extreme ultraviolet Imaging Spectrometer \citep[{EIS};][]{Culhane07},   rotation in  tornado-like prominences  has been suggested from  Dopplergrams obtained  in  the 195 \AA\ and 185 \AA\  lines     \citep{Su2014,Levens2015}.  {  These two lines are formed at coronal temperatures  ($\log T >$ 6), and thus these measurements concerned mainly the envelope or prominence-corona transition region. It is therefore  important to also
analyse the behaviour of the cool plasma inside tornadoes.}

 { Spectro-polarimetric} observations of tornadoes are rare because of the  long exposure  time needed to scan the whole structure   with a slit  (around  20 min to 1 hour). This problem is exacerbated with lines formed at low temperature. 
 Tornadoes have been observed in optical  wavelength range in  \ion{He}{i} 10830 \AA\ with the  {Tenerife Infrared Polarimeter} (TIP) operating at  the  {Vacuum Tower Telescope} (VTT) \citep[  e.g. ][]{Orozco2012,2016ApJ...825..119M}, in the \ion{He}{i} D3 line by  the {Télescope Héliographique pour l'Etude du Magnétisme et des Instabilités Solaires}  (THEMIS, the French telescope  in the Canary Islands),
  and  in H$\alpha$ at the  Meudon solar tower \citep{Schmieder2017}.   
 The cadence of the acquisition of the data at the VTT and in THEMIS is too slow to follow the fast evolution of the Dopplershift pattern in the whole structure. \cite{2016ApJ...825..119M} used the scanning mode
and obtained four consecutive spectro-polarimetric scans in four
hours. { They could not find any coherent behaviour
between the two successive scans} and concluded that if rotation
exists it must be intermittent and last less than one hour.
{  This can be explained by the paper of 
\citet{Schmieder2017}. These authors   reported observations of a tornado obtained with the {Multi-channel Subtractive Double Pass spectrograph} (MSDP) operating at the Meudon solar tower.  The MSDP has the capability to observe a FOV in nine wavelengths along the H$\alpha$ profile~simultaneously. With this instrument, Dopplershift  maps  have been  registered with a cadence of two per  minute, making it possible to   follow the  fast evolution of the Dopplershift pattern. Therefore the  authors  were able to show  that large blueshift cells become large redshift cells in a quasi periodic  manner. }

{ Using data provided by  the Hinode/SOT spectrometer working in the H$\alpha$ line   and by  the  {Interface Region Imaging Spectrograph}  \citep[IRIS;][]{DePontieu2014}  working in the doublet of   \ion{Mg}{II}, \citet{Kucera2018}   reported  observations of tornadoes. These authors also found that there
were Doppler shifts at certain times and locations. The Doppler shifts were however transitory and localised and  could 
be the result of oscillations or motions of particular
features along the line of sight.
 On the other hand, \citet{Yang2018} analysed two tornadoes in \ion{Mg}{ii}  lines with a very narrow FOV and observed  blueshifts and redshifts adjacent to each other  over a timescale of two and a half hours. They interpreted this as evidence that the tornadoes were rotating. The narrow FOV however covered only a part of the tornado, and it is not clear whether or not   this FOV  was representative of all the structure motions.}

On the theoretical side, \citet{Luna2015} presented an MHD model based on a vertical cylinder with a vertical field along its central axis, and a  { helical}  field at the periphery; it is difficult to model such a structure and obtain the corresponding Stokes parameters.  { However,  polarimetric measurements from THEMIS did not indicate the existence of vertical structures.}
 THEMIS has been observing prominences during several international campaigns with the {\it MulTi-Raies} (MTR) mode \citep{Lopez2000}. More than 200 prominences have been observed in the \ion{He}{i} D3 line. Both statistics \citep{Lopez2015} and  case studies have been published \citep{Schmieder2013,Schmieder2014}.  The main result was that the magnetic field is mainly horizontal in prominences.  Recently, the magnetic field of  several tornado-like structures has been analysed    and {  the histograms  of the inclination of the magnetic field showed a dominant horizontal component  of the field and two secondary peaks which have been interpreted up to now as the presence of microturbulence and  not a vertical magnetic field \citep{Schmieder2015,Levens2016b}.}

 Several  authors consider tornadoes to be the legs of prominences   \citep{Wedemeyer2013,Levens2016a}. Many static models consider legs or barbs of prominences  to be the accumulation of dips \citep{Aulanier1998,Dudik2008,Heinzel1999,Mackay2010, Gunar2018}. Simulations of filament  formation  by condensation  confirm that  cool plasma  would be located in the dips of magnetic field lines  \citep{Antiochos1999,Karpen2001,Luna12,Xia2014}.

Waves and oscillations are very common in solar prominences and have been reported for many years \citep[see review by][]{arregui2018}. Several tornado studies interpreted the Doppler signatures and the apparent motions as oscillations instead of true rotations. \citet{Panasenco2014} explained the motions detected in AIA as a combination of horizontal oscillations and counterstreaming flows. These motions give the impression of apparent rotation in the plane-of-the-sky (POS). \citet{Poedts2015} observed a tornado during the eruption of the prominence. The authors suggested that the {observed motions could be explained} as kink oscillations in a vertical tube. In the model of these latter authors, the tube is the heavy prominence mass suspended in the horizontal dips of a quadrupolar magnetic structure. \citet{Schmieder2017} reported Doppler measurements in a tornado structure and concluded that the observed {motions resulted from} the combination of horizontal oscillations with the global motion of the prominence associated with a parasitic polarity close to the prominence footpoint. In all cases the motions can be associated with quasi-hourly oscillations that are very common in solar prominences \citep[see, e.g.][]{Luna2018}. { In the current work, we revisit the work of \citet{Schmieder2017} in an attempt to disentangle the proper motions of the filament from the periodic motions associated with the oscillations.}

In this paper we select a tornado observed  in the 193 \AA\ filter of SDO/AIA on September 24, 2013. The rotation of this  prominence  can be visualised in  the SDO/AIA  193 \AA\ 
 movies (available on \url{https://helioviewer.org}).  
We  revisit the  observations  of the prominence obtained   at the Meudon solar tower  with the MSDP  and already analysed in  Paper I   \citep{Schmieder2017}.  {Section~\ref{sec:data} presents the data}

 and Sect.~\ref{sec:osc} {introduces} the Scargle  method  compared to the cosine fit to analyse the oscillations. We  derive the   characteristics of the oscillations: period, amplitude, phase, and propagation.   Finally  we discuss the possible modes of the  propagating waves (Section 4).{ We  find two large areas in the prominence with standing waves compatible with gravito-acoustic oscillations. In addition, a third region shows propagating waves emanating from one of the oscillating regions. These waves are consistent with magnetosonic modes propagating in the vertical direction.} }

\section{H$\alpha$ data}\label{sec:data}
The  present tornado-like structure was observed as a filament a few days before in the survey image of Meudon \citep{Schmieder2014}.

\subsection{MSDP observations}
Time sequences of H$\alpha$ MSDP observations obtained at the Meudon Solar Tower on September 24, 2013, have already been analysed  for different purposes \citep{Schmieder2014,2015ApJ...800L..13H}. In these two former papers only the quiet part of the prominence located in the northern part of the FOV was under study; this was also in the FOV of IRIS.
 The  southern part of the prominence  was analysed later and classified as an AIA tornado  (Paper I).
 Here we focus our study on the H$\alpha$ data of  the tornado structure located in the south and   indicated by a  white box in Figure \ref{MSDP}.   Two sequences  of observations  with a cadence of two images per minute were obtained   and last  approximatively 1 hour each: from 12:06 UT to 12:53 UT and from 13:04 UT to 14:02 UT. 

\subsection{Dopplershifts}
MSDP observations  consist of nine  images obtained  in nine channels of 465 arcsec $\times$ 60 arcsec. The wavelength step corresponding to the same solar point in two consecutive channels is 0.3 \AA. Calibrations and line profiles were obtained from the usual MSDP software \citep{Mein1991}.

\begin{figure}[!ht]
   \centering
    \includegraphics[width=0.5\textwidth]{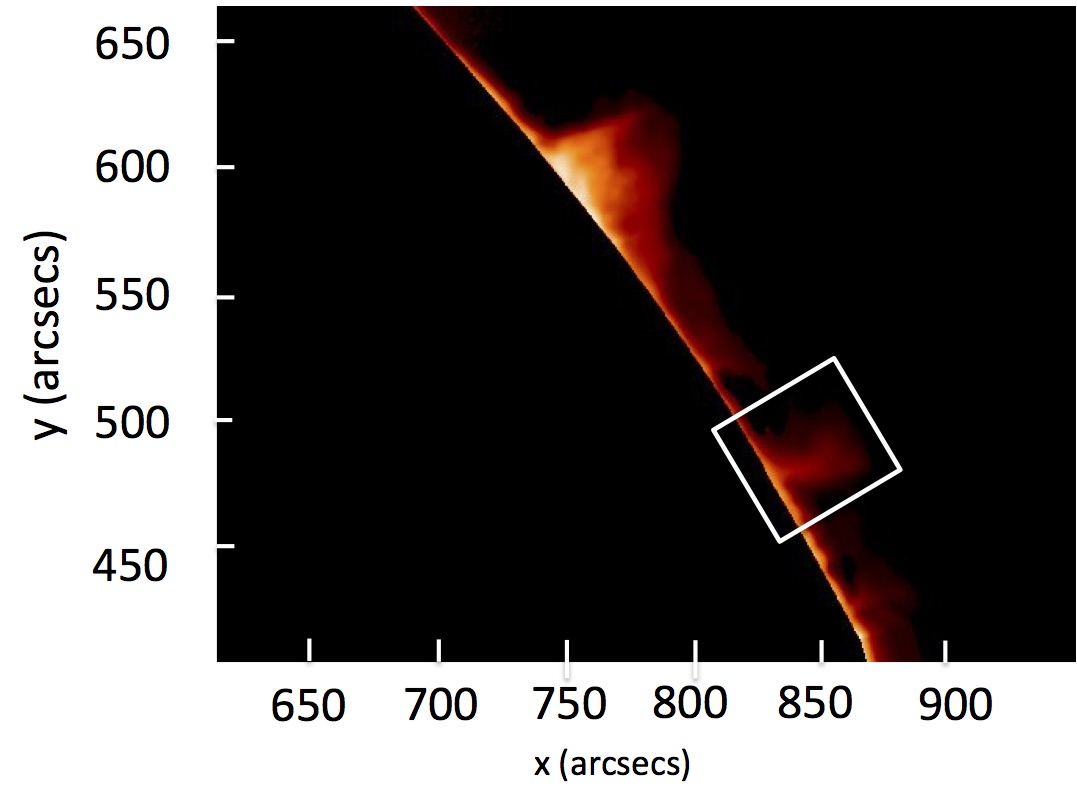}
     \includegraphics[width=0.5\textwidth]{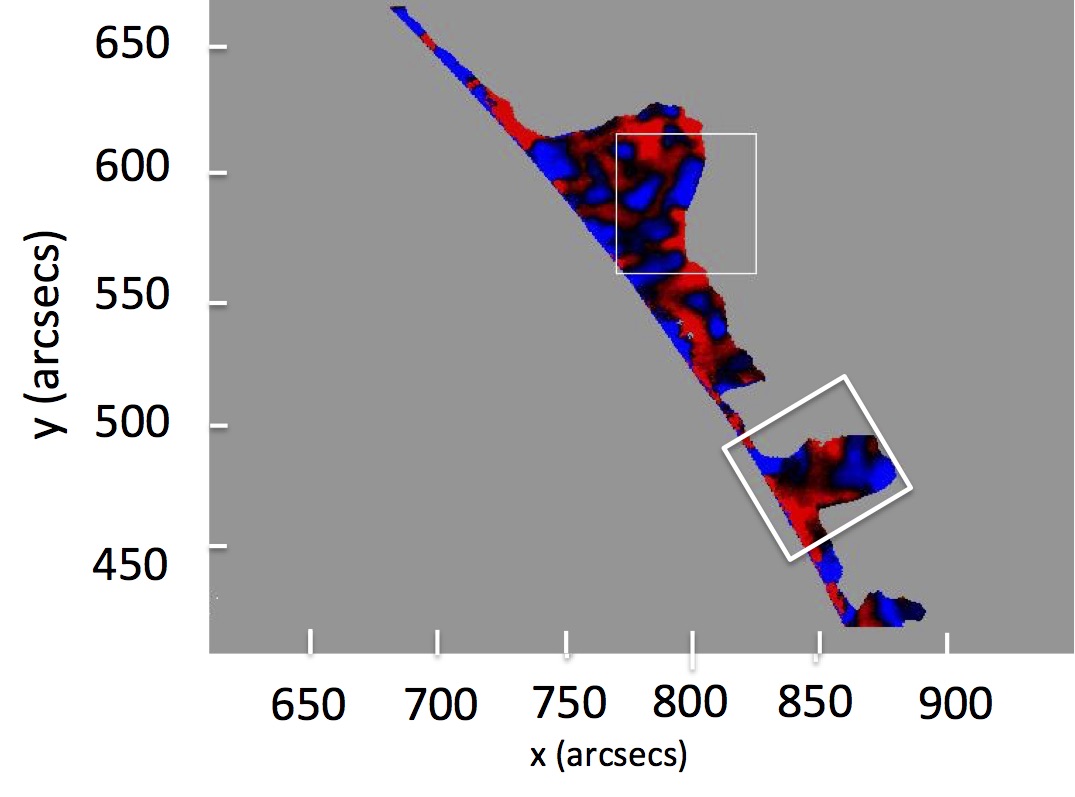}
      \includegraphics[width=0.5\textwidth]{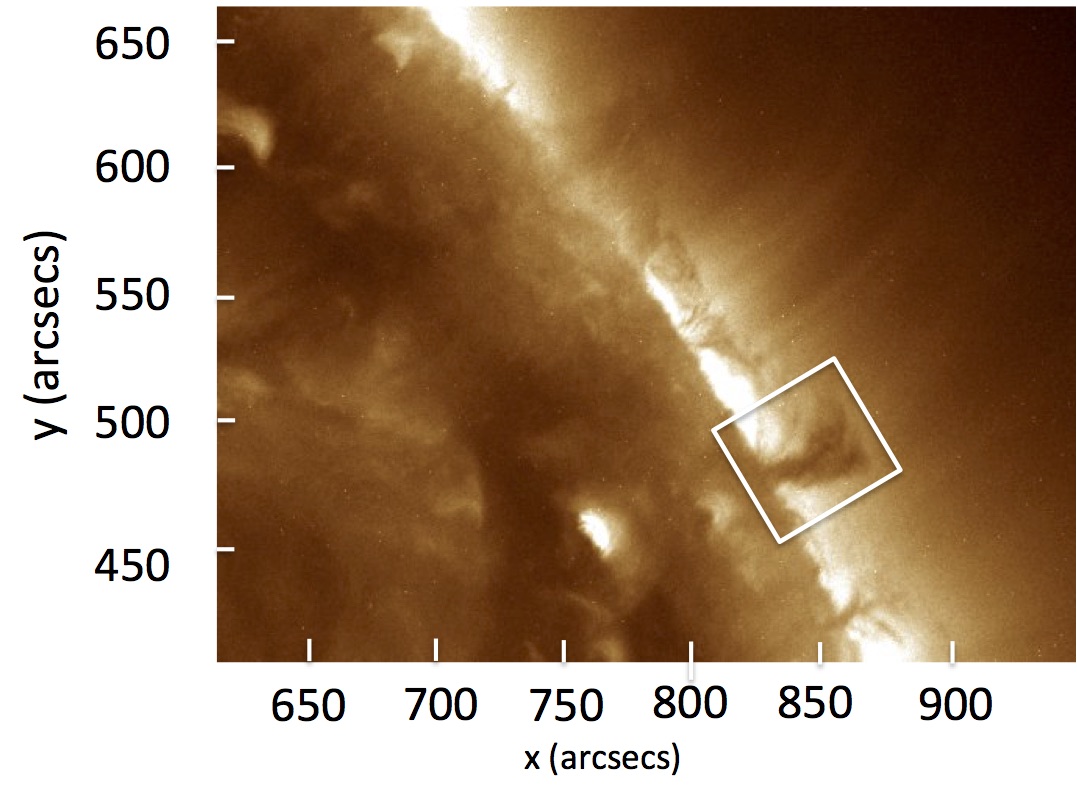}
    \caption{Prominence observed on September 24, 2013,  at 12:22 UT  with the MSDP operating on the solar tower in Meudon and  with SDO/AIA.  { Top} panel: H$\alpha$ intensity, middle panel:  H$\alpha$ Dopplershift map, { bottom} panel: AIA 193 \AA\ image. 
    The white box indicates the `tornado' (approximate FOV  of Figure \ref{maps}). }
   \label{MSDP}
   \end{figure}

The zero Dopplershift is determined using a mean value across the full prominence available in the FOV, including the upper part of the observed region (Figure \ref{MSDP}). Indeed, taking average values across the solar disc would not take into account, in particular, the departures between telluric line effects on absorption and emission profiles.

{ The scattering effects are corrected as follows. The intensity of prominence profiles is negligible in the extreme channels 1 and 9 of the MSDP. These channels can be used in each pixel of the prominence to estimate the far wings of the scattered light coming from the solar disc. A simple normalization of the disc spectrum then allows to estimate the full scattered light profile.
Because the limb is not parallel to the direction of dispersion in the spectrograph, some additional uncorrected scattered light may occur in points close to the limb, but this should not affect the computation of oscillations appreciably.

To estimate error bars for velocities, we can distinguish two kinds of errors. The first one is the data noise from CCD pixels. In Paper I, we obtained estimates from 0.2 to 0.3 km/s by using noise effects in intensity measurements and departures from velocities observed in neighbouring points. The second one concerns errors due to interpolation of line profiles. They depend mainly on slow drifts of the prominence in the FOV during time sequences, and result in stochastic wavelength drifts in each solar point. A comparison between cubic and Gaussian interpolations was used in paper I to get estimates. The result was 0.3 km/s.
Error bars due to both kinds of errors can be estimated at $\pm$ 0.5 km/s.}

\section{Oscillations} \label{sec:osc}
{ We focus the study on the Dopplershifts computed in the tornado shown in the white box in Figure 1.} In Paper I each sequence was  considered independently because of the gap of 15 minutes between them. The periodicity of the oscillations that were reported by  looking at  the variation of the Dopplershift in individual  pixels  was  close to the time interval of each sequence (around 60 minutes). Here we  add the two sequences using  a 2D Fourier correlation function and obtain one data cube of Dopplershift maps  for  a time-span of two hours.
Based on the oscillatory behaviour of each sequence taken  individually, the use of Scargle  periodograms, which compute the period of oscillations in the case of unevenly spaced data with gaps, is appropriate.

\begin{figure*}[!ht]
\centering
\includegraphics[width=0.75\textwidth]{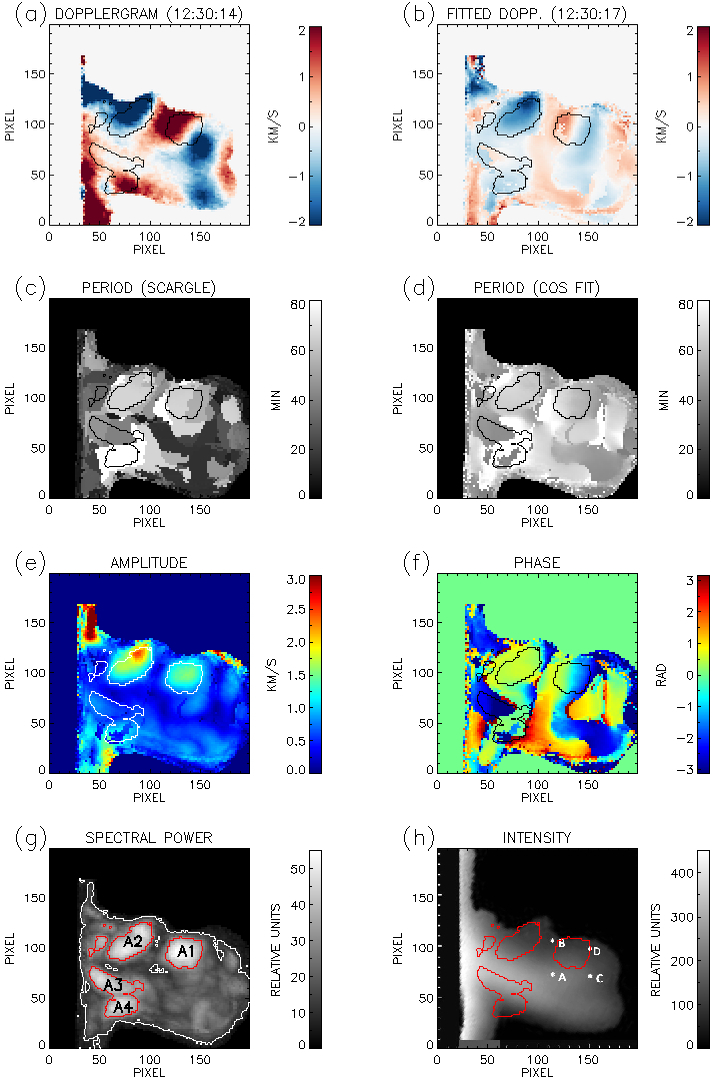}
\caption{Oscillatory features of the tornado prominence (the FOV is shown in Figure \ref{MSDP} inside the white box). (a) Snapshot of the Dopplergram movie at 12:30 UT. (b) Example of  the fitted Dopplergram at 12:30 UT. (c) Period value from periodogram analysis. (d) Period value from cosine curve approximation. (e) Amplitude value from cosine curve approximation. (f) Phase from cosine curve approximation (note that phase -$\pi$ is equal to $\pi$ radians); (g) Spectral power from periodogram analysis; white line limits areas with FAP=1\%,  red line limits areas with a spectral power higher than 32, which is also overplotted in all panels with black, white, or red lines. A1, A2, and A3 indicate areas where the mean oscillatory parameters presented in Table \ref{tab:par} are calculated. (h) H$\alpha$ intensity image with the positions of selected pixels taken as samples of Doppler signal shown in Figure \ref{dopp}. Colour-bars give units of distribution of parameters. We note that 1 pixel = 0.25 arcsec.  The temporal evolution of the Doppler maps in panels (a) and (b) is available in a movie online.
}
\label{maps}
\end{figure*}
  \subsection{Method}
\begin{figure}[!ht]
\centering
\includegraphics[width=\columnwidth]{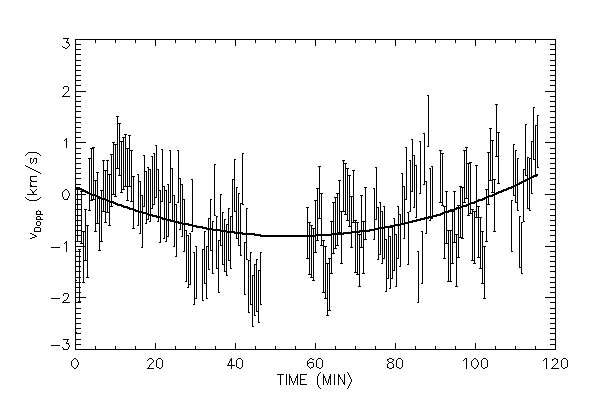}
\includegraphics[width=\columnwidth]{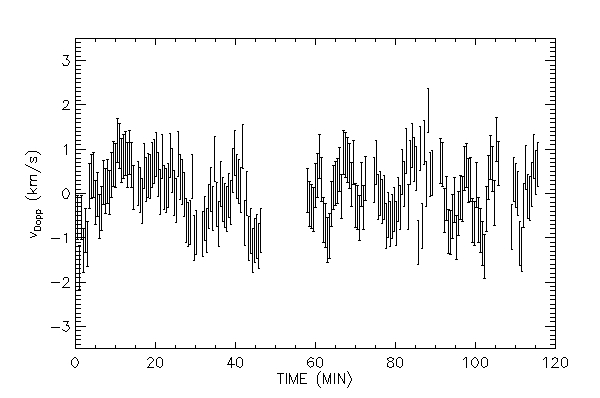}
\caption{{Example of Doppler velocity signal for pixel A with fitted parabola (top). The signal with subtracted parabola is presented in the bottom panel. { The error bar is $\pm$ 0.5 $\kms$. The gap in the data is between the two sequences observed by the MSDP.}}}
\label{parab}
\end{figure}

\begin{figure*}[!ht]
\centering
\includegraphics[width=0.9\textwidth]{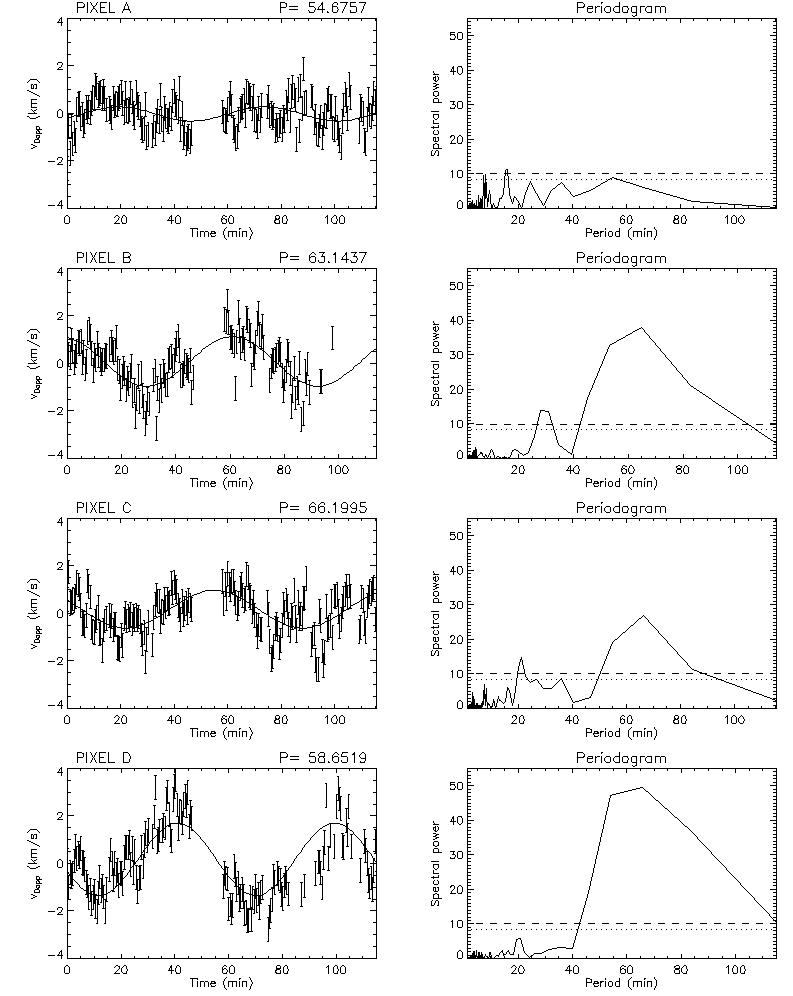}
\caption{(left) Doppler signal for  the selected pixels marked in  Fig. \ref{maps}(h) and  (right) corresponding Scargle periodograms of the signal. (+) symbols correspond to data, solid line to a cosine fit. Dashed line represents FAP=1\%, dotted line represents FAP=5\%. On the left panels, the P value gives the period of the fitted cosine curve in minutes. { The error bar is equal $\pm$ 0.5 $\kms$.}}
\label{dopp}
\end{figure*}

From the series of Dopplergrams (see Figure \ref{maps}a) we calculated the evolution of the Doppler velocity for each pixel in the field of view (FOV) independently. { The top panel of  Figure \ref{parab} shows the Doppler velocity in one pixel example. We fit each individual pixel signal by a {cosine curve}, plus a second-order polynomial function to detrend the proper motions of the prominence. The bottom panel of  Figure \ref{parab} presents the detrended signal, clearly showing the oscillatory behaviour. This trend function filters out motions associated with oscillation periods longer than the  sequence duration  and reduces instrumental effects like thermal dilatation of the spectrograph devices.}
  { Figure \ref{dopp} shows the detrended Doppler velocities in four selected pixels and their respective periodograms.}
    {  The selected pixels are shown in  Figure \ref{maps}h. We verified that the detrending has no substantial influence throughout the FOV.}

Subsequently, for all pixels, we calculate  the Scargle periodogram \citep{scargle82} which allows the identification of periodic signals in unevenly spaced data with gaps. 
Another method to deal with this kind of data 
is  the discrete Fourier transform \citep[DFT;][]{1985kurtz, 1975Deeming}. A DFT however has two disadvantages: high noise in a periodogram and spectral leakage. Scargle's formula suppresses these effects by introducing a new definition of the periodogram:
$$ P_X(\omega) = \frac{1}{2}  \left\lbrace  \frac{ \left[ \sum_j X_j \cos \omega (t_j -  \tau) \right]^2 }  { \sum_j \cos^2 \omega (t_j - \tau)}
        +                         \frac{ \left[ \sum_j X_j \sin \omega (t_j -  \tau) \right]^2 }  { \sum_j \sin^2 \omega (t_j - \tau)}   \right\rbrace ,     $$
where $X_j$ are data points observed at times $t_j$ for  ${\{ j=1,2,\ldots,N_0\} }$,  $\tau$ is defined by
$$ \tan(2\omega\tau) =  \left( \sum_j \sin 2 \omega t_j \right) / \left( \sum_j \cos 2 \omega t_j \right),  $$
and $\omega$ is calculated for the so-called natural set of frequencies:
$$ \omega_n = 2\pi n / T \quad  \{n = -N_0/2, \ldots, +N_0/2\}.$$ Additionally \cite{scargle82} defines the false-alarm probability (FAP):
$$ P(\mathrm{FAP}) = - \ln\left[ 1 - (1-\mathrm{FAP})^{1/N}\right],  $$ 
 where $N=N_0/2$. With this definition, $P(\mathrm{FAP})$ gives a power level of such a value that  the periodogram peak above $P(\mathrm{FAP})$ is caused by pure noise only with a probability equal to the FAP. The
FAP may be interpreted as a confidence level equal to ${1-\mathrm{FAP}}$.

We constructed a map of periods corresponding to the maximum peak in the periodograms calculated with the  Scargle formula in the observed prominence (Fig. \ref{maps}c).
This map shows large areas with the same or a similar period. The range of periods is mainly between 40 and 80 minutes. 
  Figure  \ref{maps}g presents the distribution of the maximum spectral power for individual pixels, showing that in almost  the whole observed prominence, the FAP level is below $1\%$.
For all pixels we then fitted a cosine function (solid lines in {left panels of  }Fig. \ref{dopp}):
$$ A \cos\left( \frac{2\pi}{P}t+\phi \right),$$
where $A$ is amplitude, $P$ is period, $\phi$ is phase shift.
We used a procedure \texttt{mpfit}, which performs robust non-linear least squares curve fitting \citep{markwardt09}. 
We limited  the range of periods to 40--80 minutes because this range covers the most prominent periods.
From the cosine curve fitting we derived  a set of parameters: $A$, $P$, and $\phi$, as immediately above. 
The spatial distribution of these parameters is shown  respectively in  Figs.~\ref{maps}(d--f). 
Having a continuous variation of Doppler signal (from curve fitting) for all pixels, we constructed  the 2D time evolution of the fitted functions. One 2D map for  a particular time may be treated as a ``fitted'' or ``approximated'' Dopplergram. The time evolution of raw and ``fitted'' Dopplergrams is presented in the animation available in the electronic version of the manuscript. 
 
  \subsection{Results}
  {  We discuss the results presented in the seven panels of Figure ~\ref{maps} in the frame of defining the nature of the oscillations.
Panels (a) and (b) are snapshots of the movie
  showing the  raw Dopplershifts and the fitted Doppler shifts, respectively, for one time, panel (c) shows the Scargle periodogram, panel (d) shows the period from cosine fitting,  panels (e, f, g) present the amplitude, the phase, and the spectral power from the cosine fitting, and panel (h) shows the H$\alpha$ intensity.}

  \subsubsection{Periods}
  We detect oscillatory motions with   periods between 5 and 80 minutes (Figure \ref{maps}c).   
  {The   shorter periods ($5$ to $ 40$ minutes) have  a local significance level of around 99\%; they last only two to four times their periods and are visible in small areas. It would be interesting to find out where and when they occur, but this is beyond the scope of this study. 
  Here, we concentrate our work on the long   periods between 40 and 80 minutes which were already detected in Paper~I. 
 We find that the prominence does  not oscillate as a block at these long periods. 
  We identify  four main oscillatory areas  with a high spectral power
  (see Figure~\ref{maps}g and Table \ref{tab:par}). 
  We select  these four areas with high spectral power  and overplot their contours on all the panels of  Figure \ref{maps}. 
  For each area, we calculate  the mean value of  the oscillatory parameters (see Table \ref{tab:par}).  The value of the spectral power in relative units is around 40. The mean oscillation period for areas A1 and A2 is close to 65 minutes with the two methods (Scargle and cosine fit methods). For  area A3 we obtain a value of 45 minutes.
  We exclude area A4~from further analysis because of the non-uniform pattern in period, amplitude, and phase distribution, which is a hint of the bad fit of the cosine curve due to high noise. Additionally, the periods calculated using the two methods are significantly different.
  We may note that a similar period value of 65 minutes is commonly observed for longitudinal oscillations in prominences \citep{Luna2018}. In areas A1, A2, and A3, the period is nearly uniform (Figure \ref{maps}c).
  The mean amplitude of the oscillations is between 0.70~km s$^{-1}$ and 1.43~km s$^{-1}$,  values larger than the estimation of the error bars.  However,  higher  amplitude  values are present  in  the selected  areas, reaching locally 3 km s$^{-1}$  (Figure \ref{maps}e). We notice another  region with even higher amplitude located  near  the pixel coordinates (x=40, y=150),  but   the signal has    a relatively low spectral power value  and  is too small to  be treated  as  an oscillating feature.}
  \subsubsection{Phase}
    Fitted phase is presented in Figure \ref{maps}f.  In areas A2 and A3  at the bottom of the {prominence} the phase has the same value in each area. 

 Additionally, in each of these two regions the period values are very similar (see Figs. \ref{maps}c and \ref{maps}d),
 and the spectral power reaches local peaks (see Figure \ref{maps}g). 
 
 On the contrary, in 
 area A{1} the period and amplitude are similar but the phase does not have a  uniform value; it shows  a continuous transition between {-$\pi$ and 0} radians. This may be interpreted as the  passage of a wavefront of a  MHD wave \citep{Schmieder2013,Ofman2015}. {The} colour gradient gives {the} direction of the propagation of the wave in the POS. This is clearly visible in the animation of the Figure \ref{maps}b available in the electronic version of the journal.

\subsubsection{Wave propagation}

\begin{figure*}[!ht]
\centering
\includegraphics[width=\columnwidth]{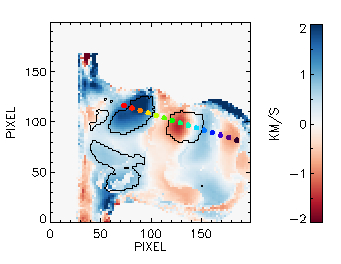}
\caption{ 
Dopplergram snapshot {rotated for convenience} with superimposed path (colour points) used to calculate time-distance diagram (see Figure \ref{tdist}). 
The colours of the points correspond to
plots in the Fig. \ref{pts}. }
\label{line}
\end{figure*}

\begin{figure*}[!ht]
\centering
\includegraphics[width=\columnwidth]{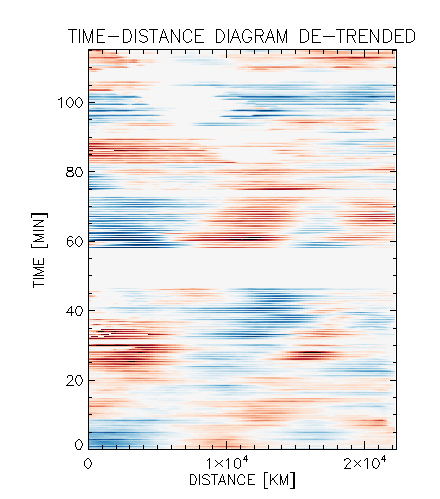}
\includegraphics[width=\columnwidth]{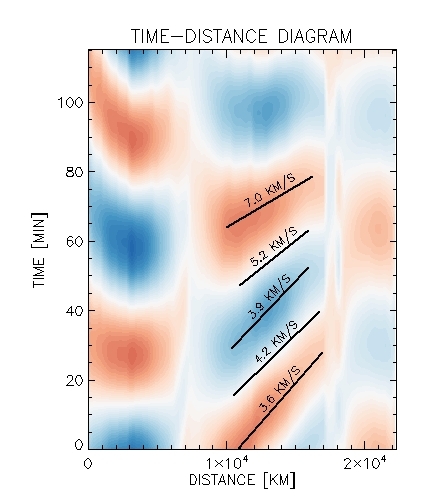}
\caption{Time-distance diagrams calculated along the path presented in  Fig. \ref{line}. Left panel: Detrended data; the white horizontal  spaces are due to the lack of data. Right panel: As in left panel but gaps are interpolated by cosine fit with the assumption of the oscillatory trend  of the data. Black lines mark positions of the same phase used to calculate the phase velocity. Numbers above lines give the calculated phase velocity {at these positions}.}
\label{tdist}
\end{figure*}

{For  a  more detailed study of  the wave propagation, we constructed  a time-distance diagram. In Figure \ref{line} we present  a Doppler snapshot with  a superimposed path, along which we calculate the time-distance diagram (Figure \ref{tdist}). }

{ Let us first comment on the accuracy of the  data. 
By comparing  Fig. 3 of  Paper I and  our Fig. \ref{line} 
we see that the direction of the path of the latter  is close to
the direction of constant wavelengths in the MSDP channels.
This implies that the mean velocity errors due to interpolation 
are similar along the path in Fig. \ref{line}, the latter being used to calculate the velocities in Fig. \ref{tdist}.}

\begin{figure}[!ht]
\centering
\includegraphics[width=\columnwidth]{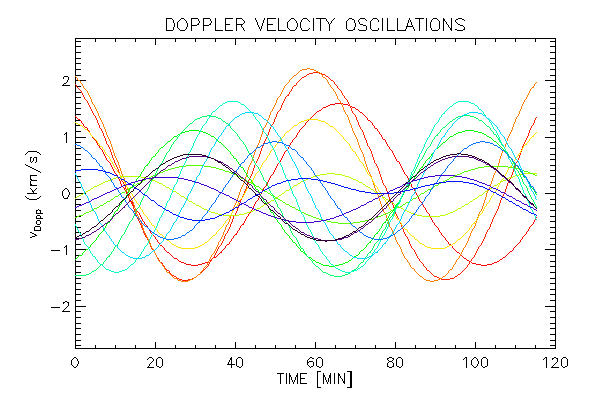}
\caption{Time variation of the Doppler velocity along the selected path. The colours correspond to positions along the path shown in Fig. \ref{line}.
}
\label{pts}
\end{figure}

{ Three different patterns  in the time-distance diagram can be distinguished in Fig. 6. The first pattern  is from 0 to $\sim$7000 km along the path from left to right  (represented by red to yellow points) and corresponds to the area A2. 
There is a small phase shift in A2 associated to a possible wave propagation in the right panel of the figure. However, looking at the left panel, we clearly see a blue and red pattern associated to standing modes. The small phase shifts could be associated to a possible artifact of the method used. For this reason we interpret the periodic motions in A2 as rigid oscillations.
 The oscillations are mainly in phase, which is visible in Fig. \ref{pts}, where variations of the Doppler velocity for  the selected pixels along the path are presented. Curves with red to yellow colours correspond to  area A2. The second feature is from $\sim$8000 to $\sim$17000 km and corresponds to area A1  and its surroundings. The oscillations are propagating along the path -- this  is visible in Fig.~\ref{pts}  from the coloured curves  (from green to purple)  which are consecutively shifted in phase. From the calculation of the position of the same phase (see Figure \ref{tdist}), we estimate  the phase velocity in this area to be 4.8 $\pm$ 1.2~km s$^{-1}$. We must keep in mind that this value is only a lower limit because of projection effects. 
AIA 193 and AIA 304 movies (available on https://helioviewer.org), where the whole prominence system was visible as a long filament on the solar disc, show that two or three days before September 24, the filament structure had a C shape and the analysed region was located in the meridional part of the system. This suggests that the analysed prominence is seen edge-on. This allows us to conclude that the real phase velocity is not much higher than calculated. The last pattern is visible from $\sim$19000 to $\sim$22000 km and can be treated as a standing wave, but it does not correspond to any area with high spectral power (see Figure \ref{maps}g).}

\begin{figure*}[!ht]
\centering
\includegraphics[width=0.95\columnwidth]{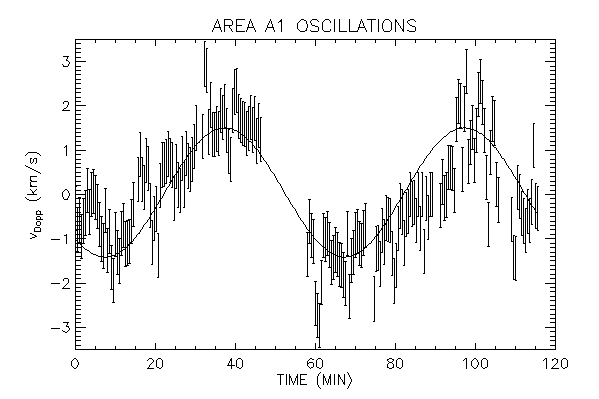}
\includegraphics[width=0.95\columnwidth]{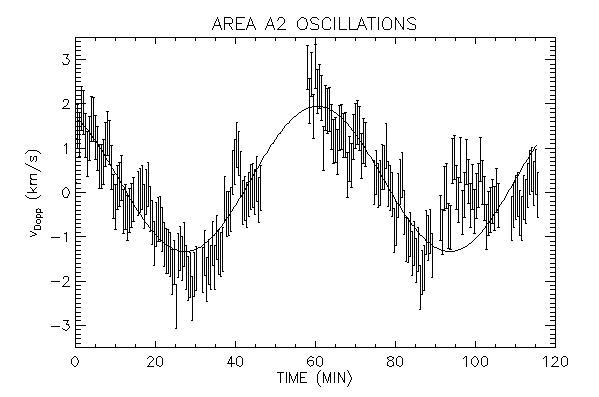}
\includegraphics[width=0.95\columnwidth]{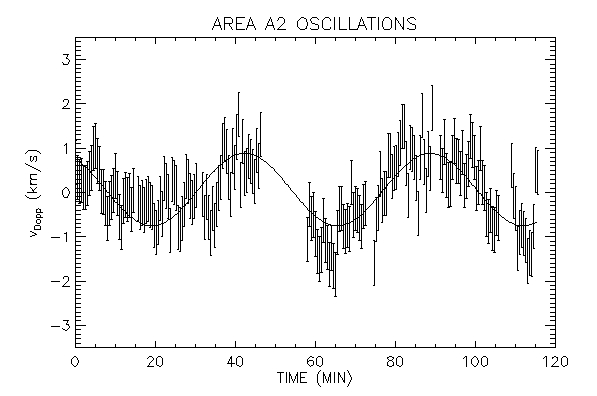}

\caption{Oscillations of the areas with maximum power spectrum. The error bars of the  data are of the order of 0.5 $\kms$; solid line - cosine fit.}
\label{aaa}
\end{figure*}

\begin{table*}[!t]
\caption{Mean oscillatory parameters calculated for areas with the highest spectral power (see Figure \ref{maps}).}
\label{tab:par}
\centering                          
\begin{tabular}{ccccc}
\hline  \hline
Area & A1 & A2 & A3 & A4\\ 
\hline 
Coordinates [pxl] &(130,90) & (74,100) & (60,70) & (70,40) \\
Spectral Power [rel.units]      & 48.50 & 41.13 & 40.45 & 39.57 \\ 

Amplitude [km s$^{-1}$]         & 1.33  & 1.43  & 0.70  & 0.32 \\ 

Period(Scargle) [min]           & 62.68 & 64.98 & 45.07 & 84.19 \\ 

Period(Cosine fit) [min]        & 62.31 & 64.44 & 45.98 & 64.20 \\ 
\hline 

\end{tabular} 
\end{table*}

\section{Discussion and conclusion}
 A tornado-like prominence (exhibiting apparent rotational motions in AIA 193~\AA\ movie) was followed for two hours at the Meudon Solar Tower with the MSDP spectrograph in H$\alpha$.  
{  A~partial  analysis of the same data  was presented in Paper I.}
 We study again the spatial variation of the  H$\alpha$ Doppler shifts in the prominence in order to see if we can detect  a rotation motion in the cool plasma (10$^4$ K). 

As in Paper I we do not find any sign of tornado-like rotation, which means that no systematic red- or blue-shift pattern is present in opposite sides of  the prominence vertical axis over large periods of time, as 
reported by \cite{Su2014}, \cite{Levens2015}, and \cite{Yang2018} from the analysis of EIS and IRIS observations of different tornadoes. 

{ In this work, we confirm the presence of oscillatory motions with periods between  40 and 80 minutes already identified in Paper I. We  extend the analysis by using the~Scargle periodogram method and find that the prominence does not oscillate at the same frequency as a block. We identify three main oscillatory areas,  A1, A2, and A3, which are disconnected.  The oscillations are  coherent inside areas A2 and A3. In area A1 the phase changes in a continuous way suggesting a propagating wave.}

{ With the MSDP we have information on the 2D oscillating pattern and we avoid detecting false oscillations caused by slit shift in slit-spectrographs \citep{Balthasar1993, zapior15}.}

{ 
The oscillations in regions A2 and A3 can be interpreted as normal modes of the prominence structure where a large part of the structure oscillates coherently. Several works considered the normal modes of prominences, modelling them as a plasma slab supported in a horizontal magnetic field \citep[e.g.,][]{joarder1992,joarder1993,oliver1993}. These latter authors found MHD modes of vibration of the prominence with periods of around one hour with horizontal velocities, in agreement with our observations. All of these modes are internal slow modes. In these modes the period is dictated essentially by the phase speed and the width of the prominence plasma slab. More recently, \citet{luna2012} and \citet{luna2012b} found that longitudinal oscillations are strongly influenced by the curvature of the magnetic field. The resulting generalised modes are called gravito-acoustic modes or pendulum modes. The authors found that in prominence conditions the gravity projected along the field lines dominates over the gas-pressure gradients as the restoring force. The period of the oscillations depends exclusively on the radius of curvature of the dipped field lines, $R$. Recent 2D and 3D numerical simulations have found that the horizontal and hourly oscillations are in agreement with the pendulum model \citep{luna2016,zhou2018} and this mode is excited after an external perturbation. We can apply the pendulum model expression \citep{luna2012,luna2012b},
\begin{equation}
P = 2\pi \sqrt{\frac{R}{g}}, 
\end{equation}
in the three regions with clear oscillations using the periods shown in Table \ref{tab:par}, that is A2-A4. In this expression, $g$ is the gravitational acceleration at the solar surface. We obtain {radii of} 104, 53, and 103$\Mm$ respectively, calculated using periods from the cosine fit of the table. In \citet{Schmieder2017} we argued that the pendulum model requires a very rigid magnetic field. However, in light of the recent realistic 3D numerical simulations by \citet{zhou2018}, we do not find a reason to reject this model.

In region A1, a {propagating} wave is observed with a phase speed between 4.9 and 7.3 $\kms$ and a period of approximately 62 minutes. This is similar {to the observation by} \citet{terradas2002} in a limb prominence. {However, these} authors found a complex pattern of wave propagation in a large portion of the prominence. This contrasts with our observation where the propagation is almost vertical. Vertical propagation of waves has already been observed by \citet{Schmieder2013}, but with much smaller periods of a few minutes associated to magnetosonic waves. The restoring force in {these} waves is the gas and magnetic pressure gradients perpendicular to the magnetic field. The authors suggested that the waves were driven by photospheric or chromospheric oscillations \citep[see also][]{Ofman2015}. \citet{mashnich2009a} and \citet{mashnich2009} studied simultaneous Doppler velocities in filaments and the photosphere underneath, finding quasi-hourly oscillations in both with good spatial correlation. This suggests that it is possible to drive quasi-hourly oscillations in prominences from below.

Alternatively to wave propagation, the observed motions can also be interpreted in terms of standing waves or oscillations. \citet{kaneko2015} found that the Alfv\'en- and/or slow-continuum in prominence structures can create the illusion of wave propagation across the magnetic field. This phenomenon can be erroneously interpreted as fast magnetosonic waves. The motion is apparent because there is no propagation of energy across magnetic surfaces. These apparent waves have been called superslow waves. \citet{luna2016} studied longitudinal oscillations associated with the gravito-acoustic modes and the authors found a similar behaviour with horizontal oscillations giving the impression of a vertical propagation. 
This phenomenon is associated with oscillations along the magnetic field where the characteristic oscillation periods depend on the parameters of each field line. In these oscillations, the characteristic periods form a continuum with smooth variations in a prominence structure. In this situation, the plasma in each field line oscillates with its own period producing phase differences that {give} the impression of propagation perpendicular to the magnetic field. \citet{raes2017} studied observationally superslow oscillations detected in a filament. Combining the observations with different models they performed seismology with superslow oscillations for the first time. They found an apparent wave velocity of 4-14 $\kms$ and periods of 59 - 76 $\mins$ similar to our observations. The authors used
the pendulum model of \citet{luna2012} in a simplified structure to show that it is possible to obtain information on magnetic field structure using superslow waves and determine the radius of curvature of field lines that support the prominence but also the position of the centre of the flux rope. However, according to the calculations of \citet{raes2017}, the phase speed decreases with time, which contradicts our findings.

In this work, we demonstrate that rotational movements in prominence tornadoes are apparent. These motions are consistent with prominence oscillations. The oscillations have a complex spatial distribution with three different regions showing a strong spectral power. In regions A2 and A3 we find standing quasi-hourly oscillations consistent with slow normal modes or with gravito-acoustic mode. In contrast, propagating waves are found in A1 travelling in the vertical direction. A possible explanation is that A2 drives the waves in A1 in some kind of wave leakage. Our observations are very similar to those of \citet{terradas2002} who found large areas in the prominence with standing waves and areas with propagating waves. The authors found that the propagating waves were generated in a narrow region with standing waves. In our region A1, the waves propagate in the vertical direction, almost perpendicular to the typically horizontal prominence magnetic field. These waves can be interpreted in terms of magnetosonic waves. 
{ New long time series observations with the MSDP and other imaging spectrographs (e.g. Fabry-P{\' e}rot type) should be done to study how frequently this kind of oscillation in prominences is detected.}}
 
\begin{acknowledgements}
We would like to thank the  team of MSDP for acquiring the observations  (Regis Le Cocguen and Daniel Crussaire). MZ is supported by the project RVO:67985815.
NL acknowledges support from STFC grant ST/P000533/1.
ML acknowledges the support by the Spanish Ministry of Economy and Competitiveness (MINECO) through projects AYA2014- 55078-P and under the 2015 Severo Ochoa Program MINECO SEV-2015-0548. ML also acknowledges the support by ISSI of the team 413 on ``Large-Amplitude Oscillations as a Probe of Quiescent and Erupting Solar Prominences''. This work was initiated during meetings of the International Team "Solving the prominence paradox" organized by NL, and support from the International Space Science Institute is gratefully acknowledged.
\end{acknowledgements}

%-------------------------------------------------------------------

\bibliographystyle{aa}
\bibliography{bibliography}

\begin{thebibliography}{56}
\expandafter\ifx\csname natexlab\endcsname\relax\def\natexlab#1{#1}\fi

\bibitem[{{Antiochos} {et~al.}(1999){Antiochos}, {MacNeice}, {Spicer}, \&
  {Klimchuk}}]{Antiochos1999}
{Antiochos}, S.~K., {MacNeice}, P.~J., {Spicer}, D.~S., \& {Klimchuk}, J.~A.
  1999, \apj, 512, 985

\bibitem[{Arregui {et~al.}(2018)Arregui, Oliver, \& Ballester}]{arregui2018}
Arregui, I., Oliver, R., \& Ballester, J.~L. 2018, Living Reviews in Solar
  Physics, 15, 3

\bibitem[{{Aulanier} \& {Demoulin}(1998)}]{Aulanier1998}
{Aulanier}, G. \& {Demoulin}, P. 1998, \aap, 329, 1125

\bibitem[{{Balthasar} {et~al.}(1993){Balthasar}, {Wiehr}, {Schleicher}, \&
  {Wohl}}]{Balthasar1993}
{Balthasar}, H., {Wiehr}, E., {Schleicher}, H., \& {Wohl}, H. 1993, \aap, 277,
  635

\bibitem[{{Berger}(2014)}]{Berger2014}
{Berger}, T. 2014, in IAU Symposium, Vol. 300, IAU Symposium, ed.
  B.~{Schmieder}, J.-M. {Malherbe}, \& S.~T. {Wu}, 15--29

\bibitem[{{Culhane} {et~al.}(2007){Culhane}, {Harra}, {James}, {Al-Janabi},
  {Bradley}, {Chaudry}, {Rees}, {Tandy}, {Thomas}, {Whillock}, {Winter},
  {Doschek}, {Korendyke}, {Brown}, {Myers}, {Mariska}, {Seely}, {Lang}, {Kent},
  {Shaughnessy}, {Young}, {Simnett}, {Castelli}, {Mahmoud}, {Mapson-Menard},
  {Probyn}, {Thomas}, {Davila}, {Dere}, {Windt}, {Shea}, {Hagood}, {Moye},
  {Hara}, {Watanabe}, {Matsuzaki}, {Kosugi}, {Hansteen}, \&
  {Wikstol}}]{Culhane07}
{Culhane}, J.~L., {Harra}, L.~K., {James}, A.~M., {et~al.} 2007, Sol. Phys.,
  243, 19

\bibitem[{{De Pontieu} {et~al.}(2014){De Pontieu}, {Title}, {Lemen}, {Kushner},
  {Akin}, {Allard}, {Berger}, {Boerner}, {Cheung}, {Chou}, {Drake}, {Duncan},
  {Freeland}, {Heyman}, {Hoffman}, {Hurlburt}, {Lindgren}, {Mathur}, {Rehse},
  {Sabolish}, {Seguin}, {Schrijver}, {Tarbell}, {W{\"u}lser}, {Wolfson},
  {Yanari}, {Mudge}, {Nguyen-Phuc}, {Timmons}, {van Bezooijen}, {Weingrod},
  {Brookner}, {Butcher}, {Dougherty}, {Eder}, {Knagenhjelm}, {Larsen},
  {Mansir}, {Phan}, {Boyle}, {Cheimets}, {DeLuca}, {Golub}, {Gates}, {Hertz},
  {McKillop}, {Park}, {Perry}, {Podgorski}, {Reeves}, {Saar}, {Testa}, {Tian},
  {Weber}, {Dunn}, {Eccles}, {Jaeggli}, {Kankelborg}, {Mashburn}, {Pust},
  {Springer}, {Carvalho}, {Kleint}, {Marmie}, {Mazmanian}, {Pereira}, {Sawyer},
  {Strong}, {Worden}, {Carlsson}, {Hansteen}, {Leenaarts}, {Wiesmann},
  {Aloise}, {Chu}, {Bush}, {Scherrer}, {Brekke}, {Martinez-Sykora}, {Lites},
  {McIntosh}, {Uitenbroek}, {Okamoto}, {Gummin}, {Auker}, {Jerram}, {Pool}, \&
  {Waltham}}]{DePontieu2014}
{De Pontieu}, B., {Title}, A.~M., {Lemen}, J.~R., {et~al.} 2014, \solphys, 289,
  2733

\bibitem[{{Deeming}(1975)}]{1975Deeming}
{Deeming}, T.~J. 1975, \apss, 36, 137

\bibitem[{{Dud{\'{\i}}k} {et~al.}(2008){Dud{\'{\i}}k}, {Aulanier}, {Schmieder},
  {Bommier}, \& {Roudier}}]{Dudik2008}
{Dud{\'{\i}}k}, J., {Aulanier}, G., {Schmieder}, B., {Bommier}, V., \&
  {Roudier}, T. 2008, \solphys, 248, 29

\bibitem[{{Dud{\'{\i}}k} {et~al.}(2012){Dud{\'{\i}}k}, {Aulanier}, {Schmieder},
  {Zapi{\'o}r}, \& {Heinzel}}]{Dudik2012}
{Dud{\'{\i}}k}, J., {Aulanier}, G., {Schmieder}, B., {Zapi{\'o}r}, M., \&
  {Heinzel}, P. 2012, \apj, 761, 9

\bibitem[{{Gun{\'a}r} {et~al.}(2018){Gun{\'a}r}, {Dud{\'{\i}}k}, {Aulanier},
  {Schmieder}, \& {Heinzel}}]{Gunar2018}
{Gun{\'a}r}, S., {Dud{\'{\i}}k}, J., {Aulanier}, G., {Schmieder}, B., \&
  {Heinzel}, P. 2018, \apj, 867, 115

\bibitem[{{Heinzel} \& {Anzer}(1999)}]{Heinzel1999}
{Heinzel}, P. \& {Anzer}, U. 1999, \solphys, 184, 103

\bibitem[{{Heinzel} {et~al.}(2015){Heinzel}, {Schmieder}, {Mein}, \&
  {Gun{\'a}r}}]{2015ApJ...800L..13H}
{Heinzel}, P., {Schmieder}, B., {Mein}, N., \& {Gun{\'a}r}, S. 2015, \apj, 800,
  L13

\bibitem[{Joarder \& Roberts(1992)}]{joarder1992}
Joarder, P.~S. \& Roberts, B. 1992, Astronomy and Astrophysics (ISSN
  0004-6361), 256, 264

\bibitem[{Joarder \& Roberts(1993)}]{joarder1993}
Joarder, P.~S. \& Roberts, B. 1993, Astronomy and Astrophysics, 277, 225

\bibitem[{Kaneko {et~al.}(2015)Kaneko, Goossens, Soler, Terradas,
  Van~Doorsselaere, Yokoyama, \& Wright}]{kaneko2015}
Kaneko, T., Goossens, M., Soler, R., {et~al.} 2015, The Astrophysical Journal,
  812, 121

\bibitem[{{Karpen} {et~al.}(2001){Karpen}, {Antiochos}, {Hohensee}, {Klimchuk},
  \& {MacNeice}}]{Karpen2001}
{Karpen}, J.~T., {Antiochos}, S.~K., {Hohensee}, M., {Klimchuk}, J.~A., \&
  {MacNeice}, P.~J. 2001, \apjl, 553, L85

\bibitem[{{Kucera} {et~al.}(2018){Kucera}, {Ofman}, \& {Tarbell}}]{Kucera2018}
{Kucera}, T.~A., {Ofman}, L., \& {Tarbell}, T.~D. 2018, \apj, 859, 121

\bibitem[{{Kurtz}(1985)}]{1985kurtz}
{Kurtz}, D.~W. 1985, \mnras, 213, 773

\bibitem[{{Lemen} {et~al.}(2012){Lemen}, {Title}, {Akin}, {Boerner}, {Chou},
  {Drake}, {Duncan}, {Edwards}, {Friedlaender}, {Heyman}, {Hurlburt}, {Katz},
  {Kushner}, {Levay}, {Lindgren}, {Mathur}, {McFeaters}, {Mitchell}, {Rehse},
  {Schrijver}, {Springer}, {Stern}, {Tarbell}, {Wuelser}, {Wolfson}, {Yanari},
  {Bookbinder}, {Cheimets}, {Caldwell}, {Deluca}, {Gates}, {Golub}, {Park},
  {Podgorski}, {Bush}, {Scherrer}, {Gummin}, {Smith}, {Auker}, {Jerram},
  {Pool}, {Soufli}, {Windt}, {Beardsley}, {Clapp}, {Lang}, \&
  {Waltham}}]{Lemen2012}
{Lemen}, J.~R., {Title}, A.~M., {Akin}, D.~J., {et~al.} 2012, \solphys, 275, 17

\bibitem[{{Levens} {et~al.}(2015){Levens}, {Labrosse}, {Fletcher}, \&
  {Schmieder}}]{Levens2015}
{Levens}, P.~J., {Labrosse}, N., {Fletcher}, L., \& {Schmieder}, B. 2015, \aap,
  582, A27

\bibitem[{{Levens} {et~al.}(2016{\natexlab{a}}){Levens}, {Schmieder},
  {Labrosse}, \& {L{\'o}pez Ariste}}]{Levens2016a}
{Levens}, P.~J., {Schmieder}, B., {Labrosse}, N., \& {L{\'o}pez Ariste}, A.
  2016{\natexlab{a}}, \apj, 818, 31

\bibitem[{{Levens} {et~al.}(2016{\natexlab{b}}){Levens}, {Schmieder},
  {L{\'o}pez Ariste}, {Labrosse}, {Dalmasse}, \& {Gelly}}]{Levens2016b}
{Levens}, P.~J., {Schmieder}, B., {L{\'o}pez Ariste}, A., {et~al.}
  2016{\natexlab{b}}, \apj, 826, 164

\bibitem[{{L{\'o}pez Ariste}(2015)}]{Lopez2015}
{L{\'o}pez Ariste}, A. 2015, in IAU Symposium, Vol. 207, IAU Symposium, ed.
  K.~N. {Nagendra}, S.~{Bagnulo}, R.~{Centeno}, \& M.~{Jes{\'u}s
  Mart{\'{\i}}nez Gonz{\'a}lez}, 275--281

\bibitem[{{L{\'o}pez Ariste} {et~al.}(2000){L{\'o}pez Ariste}, {Rayrole}, \&
  {Semel}}]{Lopez2000}
{L{\'o}pez Ariste}, A., {Rayrole}, J., \& {Semel}, M. 2000, \aaps, 142, 137

\bibitem[{Luna {et~al.}(2012)Luna, D{\'\i}az, \& Karpen}]{luna2012b}
Luna, M., D{\'\i}az, A.~J., \& Karpen, J. 2012, The Astrophysical Journal, 757,
  98

\bibitem[{Luna \& Karpen(2012)}]{luna2012}
Luna, M. \& Karpen, J. 2012, The Astrophysical Journal, 750, L1

\bibitem[{{Luna} {et~al.}(2018){Luna}, {Karpen}, {Ballester}, {Muglach},
  {Terradas}, {Kucera}, \& {Gilbert}}]{Luna2018}
{Luna}, M., {Karpen}, J., {Ballester}, J.~L., {et~al.} 2018, \apjs, 236, 35

\bibitem[{{Luna} {et~al.}(2012){Luna}, {Karpen}, \& {DeVore}}]{Luna12}
{Luna}, M., {Karpen}, J.~T., \& {DeVore}, C.~R. 2012, \apj, 746, 30

\bibitem[{Luna {et~al.}(2015)Luna, Moreno-Insertis, \& Priest}]{Luna2015}
Luna, M., Moreno-Insertis, F., \& Priest, E. 2015, The Astrophysical Journal
  Letters, 808, L23

\bibitem[{Luna {et~al.}(2016)Luna, Terradas, Khomenko, Collados, \&
  Vicente}]{luna2016}
Luna, M., Terradas, J., Khomenko, E., Collados, M., \& Vicente, A.~d. 2016, The
  Astrophysical Journal, 817, 157

\bibitem[{{Mackay} {et~al.}(2010){Mackay}, {Karpen}, {Ballester}, {Schmieder},
  \& {Aulanier}}]{Mackay2010}
{Mackay}, D.~H., {Karpen}, J.~T., {Ballester}, J.~L., {Schmieder}, B., \&
  {Aulanier}, G. 2010, \ssr, 151, 333

\bibitem[{{Markwardt}(2009)}]{markwardt09}
{Markwardt}, C.~B. 2009, in Astronomical Society of the Pacific Conference
  Series, Vol. 411, Astronomical Data Analysis Software and Systems XVIII, ed.
  D.~A. {Bohlender}, D.~{Durand}, \& P.~{Dowler}, 251

\bibitem[{{Mart{\'\i}nez Gonz{\'a}lez} {et~al.}(2016){Mart{\'\i}nez
  Gonz{\'a}lez}, {Asensio Ramos}, {Arregui}, {Collados}, {Beck}, \& {de la Cruz
  Rodr{\'\i}guez}}]{2016ApJ...825..119M}
{Mart{\'\i}nez Gonz{\'a}lez}, M.~J., {Asensio Ramos}, A., {Arregui}, I.,
  {et~al.} 2016, \apj, 825, 119

\bibitem[{Mashnich {et~al.}(2009{\natexlab{a}})Mashnich, Bashkirtsev, \&
  Khlystova}]{mashnich2009a}
Mashnich, G.~P., Bashkirtsev, V.~S., \& Khlystova, A.~I. 2009{\natexlab{a}},
  Geomagnetism and Aeronomy, 49, 891

\bibitem[{Mashnich {et~al.}(2009{\natexlab{b}})Mashnich, Bashkirtsev, \&
  Khlystova}]{mashnich2009}
Mashnich, G.~P., Bashkirtsev, V.~S., \& Khlystova, A.~I. 2009{\natexlab{b}},
  Astronomy Letters, 35, 253

\bibitem[{{Mein}(1991)}]{Mein1991}
{Mein}, P. 1991, \aap, 248, 669

\bibitem[{{Mghebrishvili} {et~al.}(2015){Mghebrishvili}, {Zaqarashvili},
  {Kukhianidze}, {Ramishvili}, {Shergelashvili}, {Veronig}, \&
  {Poedts}}]{Poedts2015}
{Mghebrishvili}, I., {Zaqarashvili}, T.~V., {Kukhianidze}, V., {et~al.} 2015,
  \apj, 810, 89

\bibitem[{{Ofman} {et~al.}(2015){Ofman}, {Knizhnik}, {Kucera}, \&
  {Schmieder}}]{Ofman2015}
{Ofman}, L., {Knizhnik}, K., {Kucera}, T., \& {Schmieder}, B. 2015, \apj, 813,
  124

\bibitem[{Oliver {et~al.}(1993)Oliver, Ballester, Hood, \& Priest}]{oliver1993}
Oliver, R., Ballester, J.~L., Hood, A.~W., \& Priest, E.~R. 1993, The
  Astrophysical Journal, 409, 809

\bibitem[{{Orozco Su{\'a}rez} {et~al.}(2012){Orozco Su{\'a}rez}, {Asensio
  Ramos}, \& {Trujillo Bueno}}]{Orozco2012}
{Orozco Su{\'a}rez}, D., {Asensio Ramos}, A., \& {Trujillo Bueno}, J. 2012,
  \apjl, 761, L25

\bibitem[{{Panasenco} {et~al.}(2014){Panasenco}, {Martin}, \&
  {Velli}}]{Panasenco2014}
{Panasenco}, O., {Martin}, S.~F., \& {Velli}, M. 2014, \solphys, 289, 603

\bibitem[{{Pesnell} {et~al.}(2012){Pesnell}, {Thompson}, \&
  {Chamberlin}}]{Pesnell2012}
{Pesnell}, W.~D., {Thompson}, B.~J., \& {Chamberlin}, P.~C. 2012, \solphys,
  275, 3

\bibitem[{Raes {et~al.}(2017)Raes, Van~Doorsselaere, Baes, \&
  Wright}]{raes2017}
Raes, J.~O., Van~Doorsselaere, T., Baes, M., \& Wright, A.~N. 2017, Astronomy
  and Astrophysics, 602, A75

\bibitem[{{Scargle}(1982)}]{scargle82}
{Scargle}, J.~D. 1982, \apj, 263, 835

\bibitem[{{Schmieder} {et~al.}(2013){Schmieder}, {Kucera}, {Knizhnik}, {Luna},
  {Lopez-Ariste}, \& {Toot}}]{Schmieder2013}
{Schmieder}, B., {Kucera}, T.~A., {Knizhnik}, K., {et~al.} 2013, \apj, 777, 108

\bibitem[{{Schmieder} {et~al.}(2015){Schmieder}, {L{\'o}pez Ariste}, {Levens},
  {Labrosse}, \& {Dalmasse}}]{Schmieder2015}
{Schmieder}, B., {L{\'o}pez Ariste}, A., {Levens}, P., {Labrosse}, N., \&
  {Dalmasse}, K. 2015, in IAU Symposium, Vol. 305, IAU Symposium, ed. K.~N.
  {Nagendra}, S.~{Bagnulo}, R.~{Centeno}, \& M.~{Jes{\'u}s Mart{\'{\i}}nez
  Gonz{\'a}lez}, 275--281

\bibitem[{{Schmieder} {et~al.}(2017){Schmieder}, {Mein}, {Mein}, {Levens},
  {Labrosse}, \& {Ofman}}]{Schmieder2017}
{Schmieder}, B., {Mein}, P., {Mein}, N., {et~al.} 2017, \aap, 597, A109

\bibitem[{{Schmieder} {et~al.}(2014){Schmieder}, {Tian}, {Kucera}, {L{\'o}pez
  Ariste}, {Mein}, {Mein}, {Dalmasse}, \& {Golub}}]{Schmieder2014}
{Schmieder}, B., {Tian}, H., {Kucera}, T., {et~al.} 2014, \aap, 569, A85

\bibitem[{{Su} {et~al.}(2014){Su}, {G{\"o}m{\"o}ry}, {Veronig}, {Temmer},
  {Wang}, {Vanninathan}, {Gan}, \& {Li}}]{Su2014}
{Su}, Y., {G{\"o}m{\"o}ry}, P., {Veronig}, A., {et~al.} 2014, \apjl, 785, L2

\bibitem[{Terradas {et~al.}(2002)Terradas, Molowny-Horas, Wiehr, Balthasar,
  Oliver, \& Ballester}]{terradas2002}
Terradas, J., Molowny-Horas, R., Wiehr, E., {et~al.} 2002, Astronomy and
  Astrophysics, 393, 637

\bibitem[{{Wedemeyer} {et~al.}(2013){Wedemeyer}, {Scullion}, {Rouppe van der
  Voort}, {Bosnjak}, \& {Antolin}}]{Wedemeyer2013}
{Wedemeyer}, S., {Scullion}, E., {Rouppe van der Voort}, L., {Bosnjak}, A., \&
  {Antolin}, P. 2013, \apj, 774, 123

\bibitem[{{Xia} {et~al.}(2014){Xia}, {Keppens}, {Antolin}, \&
  {Porth}}]{Xia2014}
{Xia}, C., {Keppens}, R., {Antolin}, P., \& {Porth}, O. 2014, \apjl, 792, L38

\bibitem[{{Yang} {et~al.}(2018){Yang}, {Tian}, {Peter}, {Su}, {Samanta},
  {Zhang}, \& {Chen}}]{Yang2018}
{Yang}, Z., {Tian}, H., {Peter}, H., {et~al.} 2018, \apj, 852, 79

\bibitem[{{Zapi{\'o}r} {et~al.}(2015){Zapi{\'o}r}, {Kotr{\v c}}, {Rudawy}, \&
  {Oliver}}]{zapior15}
{Zapi{\'o}r}, M., {Kotr{\v c}}, P., {Rudawy}, P., \& {Oliver}, R. 2015,
  \solphys, 290, 1647

\bibitem[{Zhou {et~al.}(2018)Zhou, Xia, Keppens, Fang, \& Chen}]{zhou2018}
Zhou, Y.-H., Xia, C., Keppens, R., Fang, C., \& Chen, P.~F. 2018, The
  Astrophysical Journal, 856, 179

\end{thebibliography}

\end{document}